# Stories from the Sky: Astronomy in Indigenous Knowledge

*Duane W. Hamacher*
*Nura Gili Indigenous Programs*
*University of New South Wales*
*Sydney, Australia*
*Email: d.hamacher@unsw.edu.au*

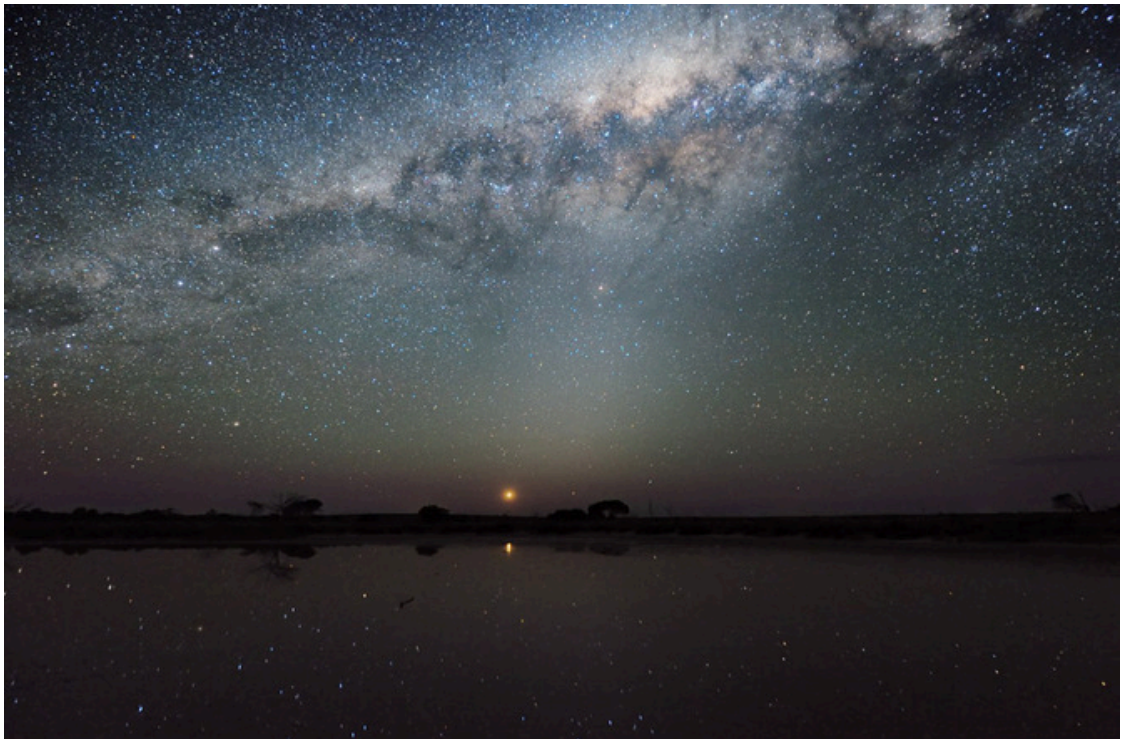

Night sky over Lake Tyrrel in Western Victoria – home of the Wergaia people. Alex Cherney, CC BY-NC-ND

*Indigenous Australian practices, developed and honed over thousands of years, weave science with storytelling. In this Indigenous science series, we'll look at different aspects of First Australians' traditional life and uncover the knowledge behind them – starting today with astronomy.*

*This article contains the names of Aboriginal people who have passed away.*

Indigenous Australians have been developing complex knowledge systems for tens of thousands of years. These knowledge systems - which seek to understand, explain, and predict nature - are passed to successive generations through oral tradition.

As Ngarinyin elder David Bungal Mowaljarlai explains: "Everything under creation […] is represented in the ground and in the sky." For this reason, astronomy plays a significant role in these traditions.

Western science and Indigenous knowledge systems both try to make sense of the world around us but tend to be conceptualised rather differently. The origin of a natural feature may be explained the same in Indigenous knowledge systems and Western science, but are couched in very different languages.

A story **recounted** by Aunty Mavis Malbunka, a custodian of the Western **Arrernte** people of the Central Desert, tells how long ago in the Dreaming, a group of women took the form of stars and danced a corroboree (ceremony) in the Milky Way.

One of the women put her baby in a wooden basket (coolamon) and placed him on the edge of the Milky Way. As the women danced, the baby slipped off and came tumbling to Earth. When the baby and coolamon fell, they hit the ground, driving the rocks upward. The coolamon covered the baby, hiding him forever, and the baby's parents – the Morning and Evening Stars – continue to search for their lost child today.

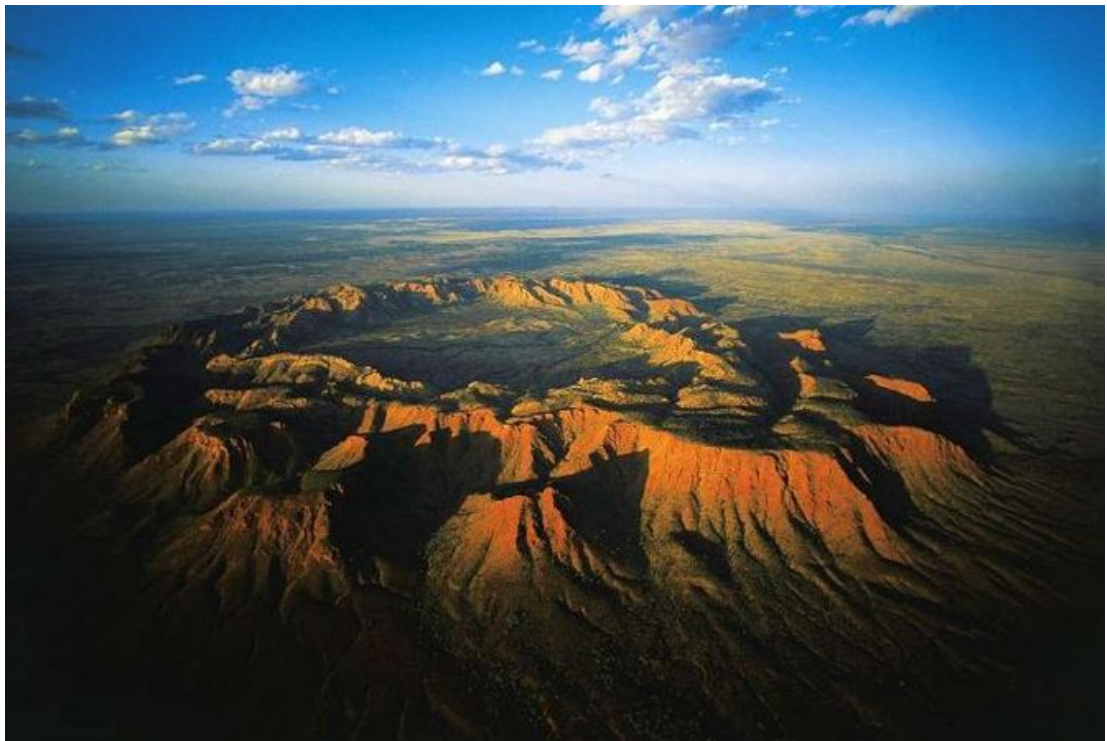
*Tnorala (Gosses Bluff crater). Dementia/Flickr, CC BY-SA*

If you look at the evening winter sky, you will see the falling coolamon in the sky, below the Milky Way, as the arch of stars in the Western constellation **Corona Australis** – the Southern Crown.

The place where the baby fell is a ring-shaped mountain range 5km wide and 150m high. The Arrernte people call it **Tnorala**. It is the remnant of a giant **crater** that formed 142 million years ago, when a comet or asteroid struck the Earth, driving the rocks upward.

**Predicting seasonal change**

When the **Pleiades** star cluster rises just before the morning sun, it signifies the start of winter to the **Pitjantjatjara** people of the Central Desert and tells them that dingoes are breeding and will soon be giving birth to pups.

The evening appearance of the celestial shark, **Baidam** traced out by the stars of the Big Dipper tells Torres Strait Islanders that they need to plant their gardens with sugarcane, sweet potato and banana.

When the nose of Baidam touches the horizon just after sunset, the shark breeding season has begun and people should stay out of the water as it is very dangerous!

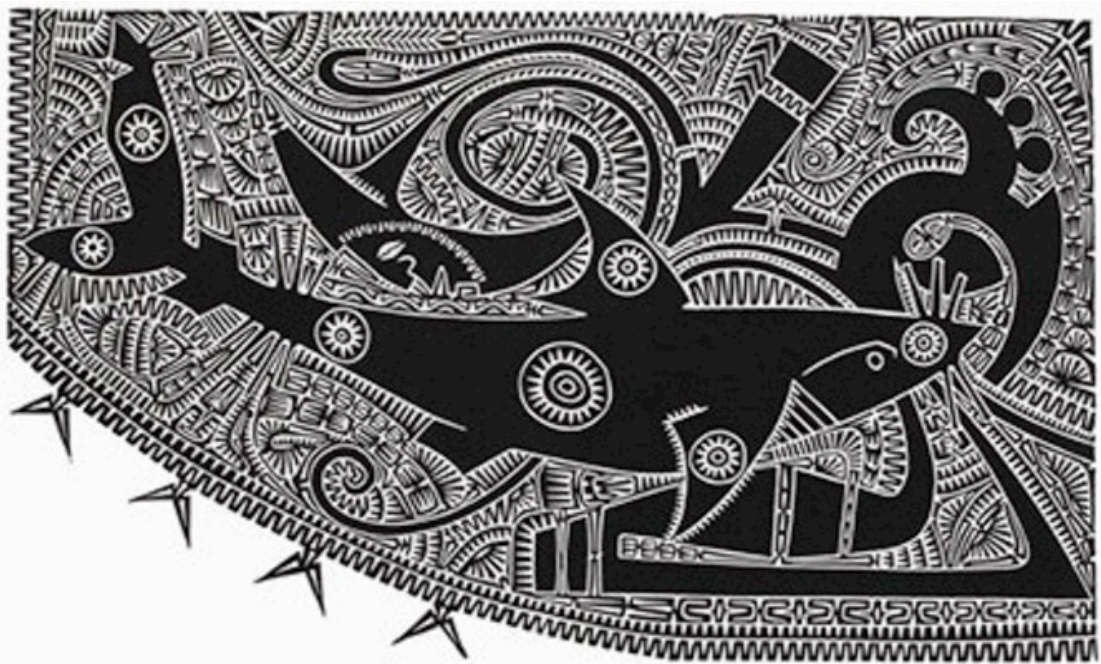

*Torres Strait Islanders use constellations, such as the shark 'Baidam' pictured here, for practical purposes. Brian Robinson*

Torres Strait Islanders' close attention to the night sky is further demonstrated in their use of stellar **scintillation** (twinkling), which enables them to determine the amount of moisture and turbulence in the atmosphere. This allows them to predict weather patterns and seasonal change. Islanders distinguish planets from stars because planets do not twinkle.

In **Wergaia** traditions of western Victoria, the people once faced a drought and food was scarce. Facing starvation, a woman named Marpeankurric set out in search of tucker for the group. After searching high and low, she found an ant nest and dug up thousands of nutritious ant larvae, called bittur.

This sustained the people through the winter drought. When she passed away, she ascended to the heavens and became the star Arcturus. When Marpeankurric rises in the evening, she tells the people when to harvest the ant larvae.

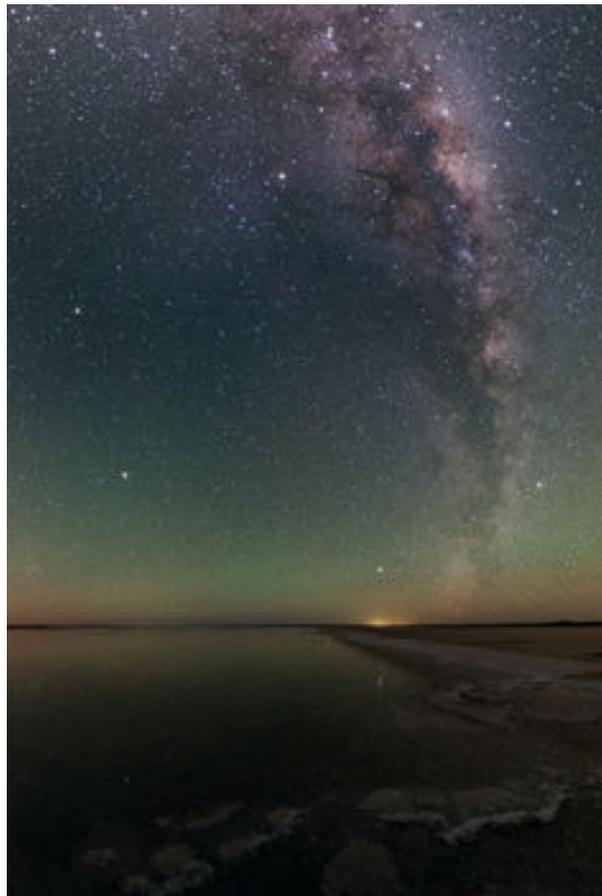

*Arcturus (Marpeankurric – on the lower left) and the Milky Way over Lake Hart. Alex Cherney*

*In each case, Indigenous astronomical knowledge was used to predict changing seasons and the availability of food sources. Behind each of these* brief accounts is a complex oral tradition that denotes a moral charter and informs sacred law.

An important thing to consider is that small changes in star positions due to stellar **proper motion** (rate of angular change in position over time) and **precession** (change in the orientation of Earth's rotational axis) means that a few thousand years ago, these sky/season relationships would have been out of sync.

This means knowledge systems had to evolve over time to accommodate a changing sky. This shows us that what we know about Indigenous astronomical knowledge today is only a tiny fraction of the total knowledge developed in Australia over the past 50,000-plus years.

**Moving forward**

As we increase our understanding of Indigenous knowledge systems, we see that Indigenous people did develop a form of **science**, which is used by Indigenous and non-Indigenous people today.

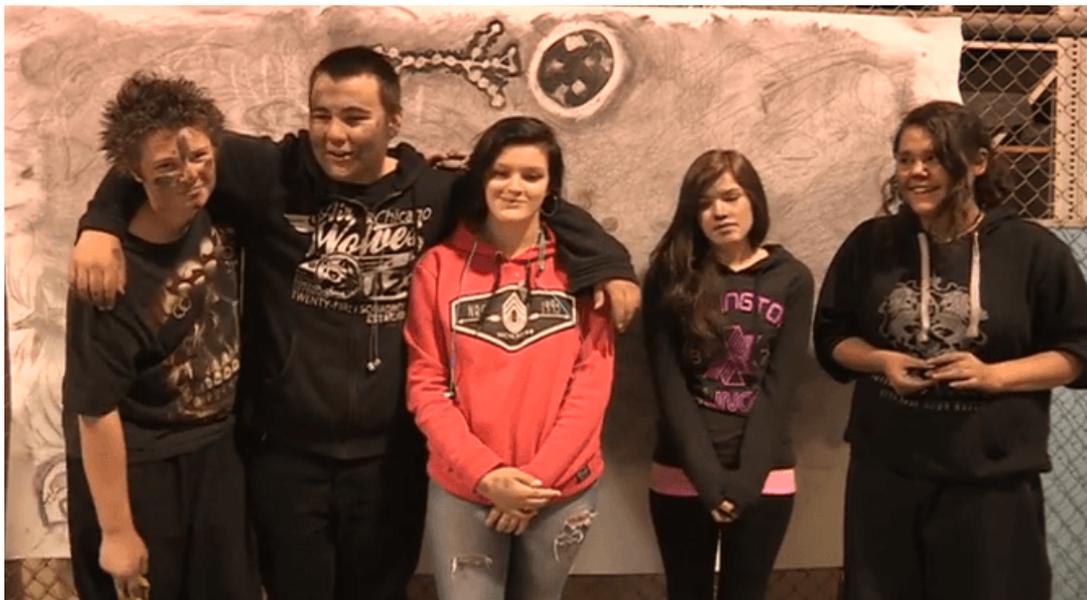
*Students in Albury/Wodonga learning about Indigenous astronomy through the Charcoal Nights initiative. Murray Arts*

Traditional fire practices are used across the country, **bush medicines** are being used to treat disease, and **astronomical knowledge** is revealing an intellectual complexity in Indigenous traditions that has gone largely unrecognised.

It is time we show our appreciation for Indigenous knowledge and celebrate the many ways we can all learn from this vast accumulation of traditional wisdom.

**References**

Hamacher, D.W. (2012). On the astronomical knowledge and traditions of Aboriginal Australians. Doctor of Philosophy thesis, Department of Indigenous Studies, Macquarie University, Sydney, Australia.